\documentclass{mem}
\usepackage{natbib}\usepackage{txfonts}\usepackage{balance}
\usepackage{graphicx}
\usepackage[a4paper,breaklinks,dvipdfm]{hyperref}
\idline{22}{19}

\begin{document}

\title{The lithium problem}%, a phenomenologist's perspective}
   \subtitle{A phenomenologist's perspective}

\author{
Fabio Iocco
}

%$^7{\rm Li}$

\institute{The Oskar Klein Center for CosmoParticle Physics, Department of Physics, Stockholm University, Albanova, SE-10691 Stockholm, Sweden}

\authorrunning{Fabio Iocco}

\titlerunning{the lithium problem}

\abstract{
Thirty years after the first observation of the $^7{\rm Li}$ isotope in the atmosphere of metal-poor halo stars,
the puzzle about its origin persists.
Do current observations still support the existence of a ``plateau'': a single value of lithium abundance, constant over several
orders of magnitude in the metallicity of the target star?
If this plateau exists, is it universal in terms of observational {\it loci} of target stars?
Is it possible to explain such observations with known astrophysical processes?
Can yet poorly explored astrophysical mechanisms explain the observations or do we need to invoke physics beyond the standard model of Cosmology and/or the standard model of Particle Physics to explain them?
Is there a $^6{\rm Li}$ problem, and is it connected to the $^7{\rm Li}$ one?
These questions have been discussed at the Paris workshop ``Lithium in the Cosmos'', and I summarize here its contents, providing an overview from the perspective of a phenomenologist.
}

\maketitle{}

\begin{flushright}
{\it -``I can't be waiting for death in a hotel room. [...] My life... my life has been spectacular.''\\
-``Now, please, take your lithium.''}\footnote{[P. Sorrentino, \href{http://www.imdb.com/title/tt0398883/}{``The consequences of love''}, Fandango productions, 2004]}
\end{flushright}

\section{A short introduction}
The seminal observation of the $^7{\rm Li}$ isotope
in low-mass turn-off stars of our galactic halo, returned a surprising result:
the lithium abundance observed in the atmosphere of the target stars -whose metallicity
spanned more than one order of magnitude in [Fe/H]- exhibited a constant behavior, \citet{spite82}.
This permitted the identification of a ``plateau'' of $^7{\rm Li}$ in turn-off,
metal-poor stars of our halo, given the very little dispersion of observations
around a single value, an observational evidence that 
was puzzling for reasons that can be summarized by the following
questions.
If the observed $^7{\rm Li}$ has been processed by the host stars, how is it possible that it exhibits such
a regular behavior over such a large interval in metallicity of the host star?
If it has not been processed by the star, the $^7{\rm Li}$ we observe is the same
that was in the atmosphere from the very beginning of the stellar formation process.
If depletion from intervening chemical evolution is negligible (as it safely turns out to be) the latter must be identified
with what is leftover from Primordial Nucleosynthesis,
but then, why is it the only element not to match the standard Primordial Nucleosynthesis
predictions made using the latest (independent) cosmological data on the baryon density?
Thirty years later, the puzzle about the origin of the $^7{\rm Li}$
observed in the atmosphere of such stars is far from being solved.
In spite of the bigger sample of stars, better quality spectra and amazing
progress in the determination of stellar atmospheres, we still struggle to understand
the origin of the observed $^7{\rm Li}$.
In these proceedings, I summarize the view that a phenomenologist (not an observer,
not an hard-core cosmologist, not a stellar theorist) has on the problem, in light
of the latest findings as discussed in the Paris workshop.
Observations are summarized in the first section, and the leading
stellar theory arguments invoked to explain the observations are shortly summarized
in the second section. In the third, current issues regarding 
Primordial (or Big-Bang) Nucleosynthesis (BBN) are briefly exposed.
A very short update on observations of $^6{\rm Li}$ is provided before finally
reporting the biggest open questions to date in the Conclusions section.

Throughout these proceedings the references to the original
literature must be limited and therefore utterly incomplete;
for each topic, I also address to the material presented
at \href{http://www.iap.fr/lithiuminthecosmos2012}{this workshop} 
with the name of the author
presenting such results: a more complete list
of references for each topic, as well as more quantitative details
can be found in their proceedings.

%These proceedings are meant to be a general overview, and shall by no means
%considered a comprehensive review.
%################
\section{Observations}
\label{sec:observ}
%################
The abundance of $^7{\rm Li}$ in stellar atmospheres is
determined through the observation of this element's doublet line
at 670.8 nm. 
In order to determine the relative abundance of $^7{\rm Li}$, 
the absorption strength of this line must be calibrated through a process that 
involves detailed knowledge of the temperature of the atmosphere, and a 
careful treatment of the latter.
For as concerning these latter issues may be ---with different groups using
different temperature scales based on different atmosphere models--- 
throughout the last thirty years they have been
fully addressed by all the different observational groups.

For some differences in the results of different groups
do arise in some cases, as thoroughly
discussed at the workshop by W.~Aoki, J.~Melendez, S.G.~Ryan, and L.~Sbordone,
they are minor with respect to the issues which are going to be faced in the following.

Issues related to the temperature scale and its effects on 
retrieving the correct atmospherical abundances 
has been thoroughly discussed in the talk by S.~G.~Ryan,
to whose proceedings I address here. Over-summarizing its conclusions, 
the solutions to the problems raised in the following discussion seems
to lie not in the atmospheric temperature scale, based on both
LTE and NLTE calculations.

In the following the $^7{\rm Li}$ abundance will be expressed in
the customary {\rm dex} unit {\rm A(Li)}=12+Log[N(Li)/N(H)].

%#########
\subsection{Halo stars}
%#########

Turn Off point (TOP), metal-poor stars in the halo of our Galaxy are selected on the basis of their kinematics,
in order to make sure they belong to the halo population and not to the younger
one present in the disk, and on their temperature $T_{eff}>$5900K.
This technique for selecting Population II stars is established and has been 
adopted throughout the last thirty years in a universal fashion.

The data presented by
W.~Aoki and L.~Sbordone
seem to converge for metallicities -2.8 $\leq$ [Fe/H] $\leq$ -2.0 :
$^7{\rm Li}$ abundance in dwarf, turn-off stars of the sample is contained,
for these metallicities, in a band 
2.1 $\leq$ ${\rm A(Li)}$ $\leq$ 2.4, \citet{aoki09}, \citet{sbordone10}.
This is the old region of the ``plateau'', which appears,
although with a somewhat increased spread around a central value
to show little sign of depletion.
The data presented by J.~Melendez, 
disagree with the latter, as he claims ``strong evidence for lithium
depletion at all metallicities'', as well as a consistent spread
of the former-thin ``plateau'', \citet{melendez10}. 
The reason for this discrepancy is yet to be understood
and will be very likely fully discussed in the forthcoming
analysis by the author.

\begin{figure*}[t!]
\resizebox{\hsize}{!}{\includegraphics[clip=true]{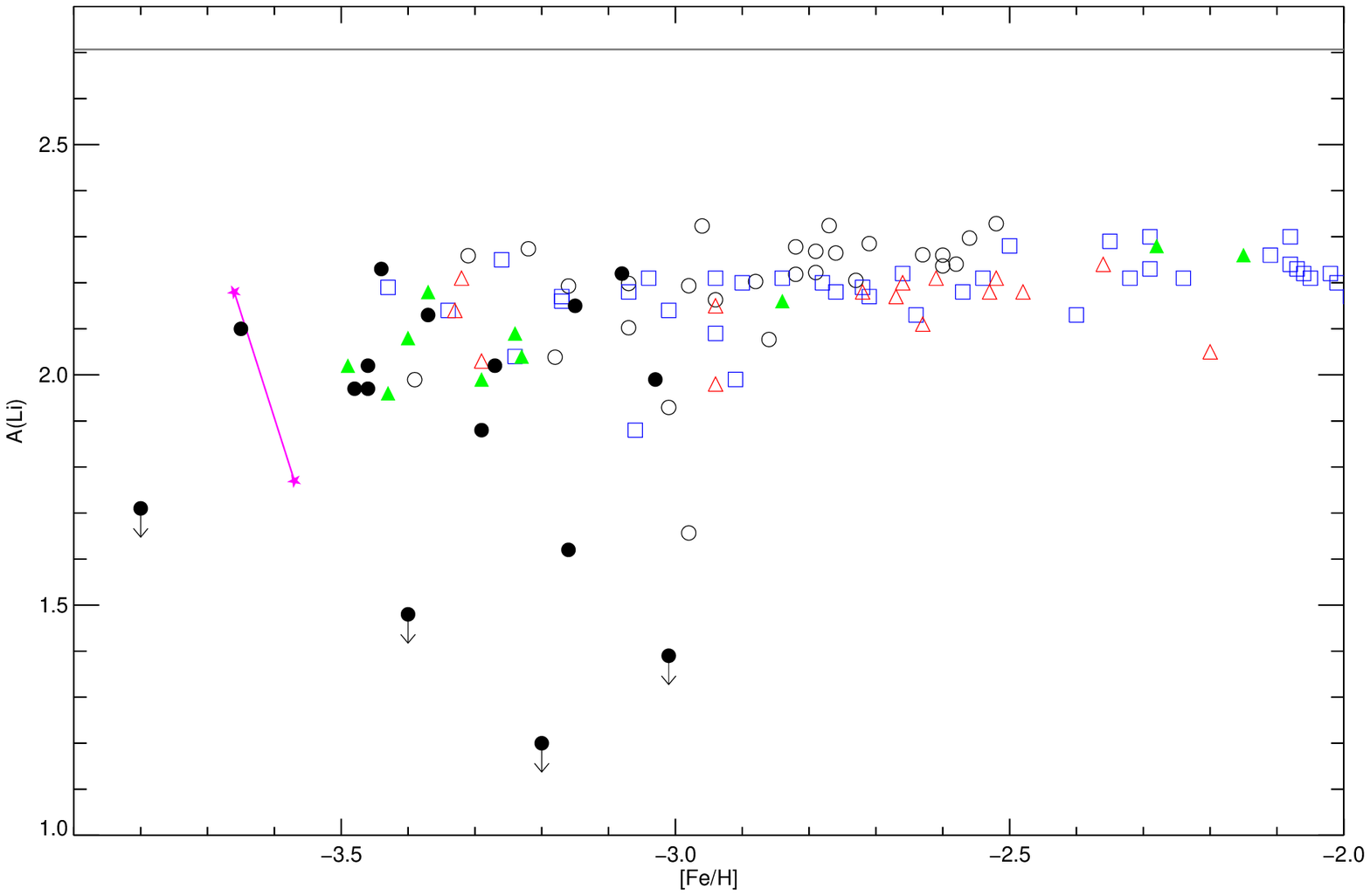}}
\caption{\footnotesize  The $^7{\rm Li}$ ``plateau'', a.~d. 2012. Courtesy of L.~Sbordone. Filled black circles, \citet{bonifacio12}; open black circles, \citet{sbordone10}; green filled triangles, \citet{aoki09}; red open triangles, \citet{hosford09}; blue open squares \citet{asplund06}; magenta stars: the two components of the binary star CS 22876-032 \citep{gonzalez08}.}
\label{fig:plateau}
\end{figure*}

Even more compelling is the evidence for a ``meltdown'' of the plateau
at metallicities below [Fe/H] $\leq$ -2.8. Such evidence appears
clearly in the data of -and has been discussed by-
W.~Aoki, J.~Melendez, and  L.~Sbordone:
whereas in no case the lithium abundance raises above the {\rm envelope} of 
{\rm A(Li)}=2.4, the scatter below such value is much more consistent, as it can clearly be
seen in the recollection of data by all groups presented by L.~Sbordone and here
reported in Figure \ref{fig:plateau}, not including the Melendez data.

The state of the art on $^7{\rm Li}$ observations in
the atmospheres of turn-off, halo-field stars can
be summarized as follows:

{\it i)} throughout the metallicity range [Fe/H] $\leq$ -2.0, the $^7{\rm Li}$
abundance {\rm A(Li)} does not raise above the ``envelope'' of {\rm A(Li)}=2.4 (see later in
Subsection \ref{sec:outliers} for exceptions, which do not affect the generality of this statement);

{\it ii)} in the metallicity range -2.8 $\leq$ [Fe/H] $\leq$ -2.0, the abundance lies
 within the interval 2.1 $\leq$ {\rm A(Li)} $\leq$ 2.4, with non-zero dispersion
even  within each single dataset, although the
degree of such dispersion varies from group's dataset to group's dataset;

{\it iii)} at metallicities [Fe/H]$\leq$-2.8, clear dispersion of several
dexes is seen below the mentioned envelope of {\rm A(Li)}=2.4.
The entity and the possible dependence of this feature is yet to be 
entirely characterized, yet it emerges clearly from the analysis
of all the group, to the point that the behavior has been named
a ``meltdown'' of the plateau.

Regarding to the ``plateau'' itself, in light of the points summarized here,
we wonder whether this term should now be used as 
referring to the envelope, rather than to a single
value, which is clearly now unsupported by observational evidence.
See more in the Conclusions.

%#########
\subsection{Globular Clusters}
%#########

Globular Clusters (GCs) are extremely valuable observation {\it loci}
in light of their special characteristics with respect to stellar formation:
stars born inside a given GC are approximately coeval,
with a uniform initial chemical signature. 
The age of stars can be determined from photometry,
therefore permitting to draw an evolutionary pattern for diffusion,
throughout the whole sequence of the Population II stars observed.
These characteristics permit to use them as extremely important
case studies, aimed at understanding the behavior of $^7{\rm Li}$ throughout
the whole evolutionary path, using the stellar population as a homogenous ensemble
of given initial metallicity.
In principle, this permits to test the lithium depletion mechanism at
different metallicities, thus ``sampling'' the X-axis of the plateau plane,
provided that the peculiar chemical history of each single cluster is 
understood, avoiding that contamination due to the local
chemical history pollute more general results.
Several individual GCs at different metallicities have been studied so far:
47 Tuc, NGC 6752, M4, M30, M92  $\Omega$ Centauri, and NGC 6397.
The range of metallicities of these objects spans from -2.8 $\leq$ [Fe/H] $\leq$ -1.1,
from the most metal poor M30, M92 up to the iron richest M4,
thus covering (although in a non-continuous fashion) a big region of the
range of halo-field stars.
The case
of the quite metal-poor NGC 6397 -[Fe/H]=-2.0- has been long debated in literature,
as summarized by A.~Korn in his review talk on GCs.
In fact, stars at different evolutionary stages exhibit a clear trend on $^7{\rm Li}$
abundance versus the effective temperature $T_{eff}$ of target stars.
The evolution of ${\rm A(Li)}$ from Turn-Off, through the 
Sub-Giant Branch and the Red Giant Branch shows a pattern which is
compatible with theoretical models including atomic
diffusion and microturbulence -see the following section \label{sec:theory}
for more details.
It is worth stressing that the $^7{\rm Li}$ abundance
observed in  turn-off stars is
compatible with that of halo-field stars observed in the halo, 
as from the previous Section, remaining
constrained within the  2.1 $\leq$ {\rm A(Li)} $\leq$ 2.4.

%#########
\subsection{Outliers in clusters and halo}
\label{sec:outliers}
%#########

On the top of this general view, whose strength relies on the
strong statistics of the data sets,
%additional single clouds loure upon the horizon:
observations of single cases of $^7{\rm Li}$-rich stars
seem to spice up some more the problem.
It is necessary to stress that the following cases
are statistically irrelevant, namely do not affect
the conclusions that one may (or may not) have reached 
over the debate of the origin of the TOP $^7{\rm Li}$ abundance.
Even the richest samples, that in the RAVE survey presented
here by G.~Ruchti, and the giants in the Milky Way Dwarf Galaxy satellites
presented by X.~Fu, do not exceed the 0.44\% of the total
sample.
In particular, a
super-$^7{\rm Li}$-rich turnoff star in NGC 6397 and one in M4 presented
respectively  by A.~Koch and L.~Monaco, 
the existence of thirteen $^7{\rm Li}$-rich giants in the Dwarf Galaxies of
the MW presented by X.~Fu, of fourteen $^7{\rm Li}$-rich field halo giants 
presented by M.~Adamow, of eight $^7{\rm Li}$-rich field
stars in the RAVE survey, and of one $^7{\rm Li}$-rich in M68, presented by
G.~Ruchti, have been discussed, \citet{koch11}, \citet{monaco12}, \citet{ruchti11}, \citet{adamow12}.
The nature of such anomalous lithium abundances -in most cases even above
the BBN predicted floor is still a puzzle,
but given the nature of the sample, much reduced with respect to the bulk of 
evidence discussed in the previous Sections- remain puzzling,
but very likely disconnected from what is typically known as the ``Lithium Problem'',
thus very likely identifying some more Lithium problems, likely of different nature.
Possible explanations proposed include ingestion of a planet or
brown dwarf because of the star's expansion, mass transfer from an AGB companion,
and self production of the $^7{\rm Li}$ the giant itself through
either cool-bottom processes, a lithium flash or a lithium circumstellar shell.
All of these explanations do however present problems, as overviewed at this
workshop by X.~Fu.

\subsection{Non-stellar abundances of lithium}
\label{sec:freespace} 

In light of what described so far, and of
the difficulties in convincing oneself on the actual
nature of the $^7{\rm Li}$ observed in the atmosphere of stars,
there is most probably very little arguing
that a measure of lithium in the most unpolluted
regions of our sky, would most probably push the needle
toward one solution or the other.
For as obvious as it seems, it is also as obviously
difficult a measure to perform, given the very low abundance
of $^7{\rm Li}$.
C.~Howk has presented at this meeting the first
observation of $^7{\rm Li}$ in a non-stellar, low-metallicity,
extra-galactic medium: 
lithium in the interstellar material of the Small Magellanic Cloud,
a Galaxy with a metallicity of one quarter that of the
Sun, [Fe/H]$\sim$-0.6 , \citet{howk12}.
The observed abundance of $^7{\rm Li}$ is
{\rm A(Li)$_{SMC}$}=2.68 $\pm$ 0.16;
which is nicely compatible with the primordial abundance,
see later section, and non compatible with the plateau envelope.
It is however to be stated that this is the first measurement of the kind,
although the region of such measurement has been chosen with
great care in order to avoid possibilities of contamination and enrichment.
This observation is clearly inconclusive with respect to the lithium problem,
but is equally clear that such technique has an enormous potential to shed
light on the problem, and observations of this type should be encouraged and
supported by the whole community.

%################
\section{Stellar Theory}
\label{sec:theory}
%################

The obvious question that arises after having overviewed 
the observations is the following:
what is the origin of the lithium in the stellar atmospheres?
This most probably constitutes the bulk of what is commonly named the ``Lithium Problem''.

On one hand one would {\it not} expect that elements in the stellar atmospheres
resemble the original composition of the star: internal mechanisms as well as 
external ones can easily enrich or deplete the star of elements heavier than Helium and
Hydrogen.
On the other hand, it is clear that the seemingly bizarre behavior of Lithium 
is not easily explained in terms of known astrophysical processes, and the 
existence of a single value, i.e. the ``plateau'' has long compelled people toward
the identification of that plateau value as the one in the original composition
of the star. Such puzzle remains even if one drops the paradigm of a single plateau value 
(in light of the previously discussed observations)
and decides to adopt the envelope value as a new plateau.

$^7{\rm Li}$ is a very fragile element, and proton capture easily destroys it at
temperatures above $\sim$2.5$\times$10$^6$K; the atmospheres of the stars are much colder than this
but convection can bring the lithium down from the cold atmosphere to hotter regions.
However, for the stars which are customarily chosen -Turn Off Point,
warm, metal-poor stars whose mass is estimated to be less than one solar mass-
the convective region below the atmosphere is not deep, and it can therefore not
drag lithium down to regions with temperatures in excess of the mentioned threshold.

In addition to this paradigmatic and yet simplistic view of the problem,
one element must be added: atomic diffusion, which slowly brings heavier elements down
in the stellar bosom.
As written by G.~Michaud in 1984, and reminded at this meeting by M. Spite:
``{\it atomic diffusion can not be turned off, it can only be
rendered inefficient by sufficient mass motion  either due to meridional circulation or turbulence}''.
Nonetheless, this essential element has been almost thoroughly omitted
for several years. 
As a matter of fact, whereas atomic diffusion does deplete the original ZAMS $^7{\rm Li}$ abundance
throughout the Main Sequence, it can not reproduce the observed abundances at the turn off point,
if the ZAMS is evolved starting with a BBN abundance.
See e.g. M.~Spite and A.~Korn in these proceedings series.

On the top of this, turbulence in the stellar
atmosphere ---caused by differential rotation, meridional circulation, gravity
wave excitation and shear---
mitigate the effects of atomic diffusion, stirring the material and hampering its fall.
All these effects are extremely challenging to model in a self-consistent way, from first principles,
in order to properly include it stellar evolution models. As discussed by
C.~Charbonnel, this is as of today yet challenging, although a better handle on the theory
can be obtained by tuning the physics on the observations of metal-richer stars than the
ones that are used for the determination of lithium abundances, \citet{talon04}.

Therefore, when dealt with, such effects are generally referred to as of an effective
turbulence (or {\it micro}turbulence), which is in some cases taken 
into account in chemical evolution models of the star.
It is however added as a parameter, whose value must be tuned rather than derived from first principles,
as mentioned above; O.~Richard has thoroughly discussed this, 
and the different types of models adopted in \citet{richard05}.
In particular, this technique is applied in order 
to reproduce the actually {\it observed}
pattern of $^7{\rm Li}$ -starting from the Standard BBN value
at Zero Age Main Sequence- in Globular Clusters.
This technique is successful in some cases, see e.g.
NGC 6397, the most debated example, \citet{korn06}, \citet{nordlander12};
however, the very same parameters adopted for the reproduction of
the NGC 6397 features  fail to work in reconciling the
observed evolutionary pattern, with the given initial conditions
for other GCs, i.e. M30 and M4.
This has been explicitly mentioned by A.~Mucciarelli,
and he has stressed that
 such ``ad hoc'' turbulence models (namely the same models
 with the very same parameters) which
successfully reproduce the whole evolutionary trend of NGC 6397,
fail to capture the subgiant enhanced abundance when
applied to M30 observations, conversely those set of 
parameters able to explain the dredge-up predict a
depletion not able to fill the gap between the BBN
and the TOP abundance.\\
It is extremely interesting to notice the following: 
Mucciarelli et al can {\it not} reproduce the global trend of Li evolution in the 
open cluster M4 using atomic diffusion and microturbulence
with the same parameters used by Korn et al. for the analysis of
NGC 6397.
However, they equally fail to reproduce the lithium trend if they adopt atomic diffusion only.
If needed, this is clear and obvious indication that atomic diffusion can not be neglected
when modeling the evolution of lithium in the atmospheres of target stars, nor is the only answer to the problem.
%################
\section{Primordial Nucleosynthesis}
\label{sec:bbn}
%################
In order to understand the origin of the
lithium observed in stellar atmospheres one piece of the puzzle must be clear:
the initial conditions expected at the best of our knowledge.
In the current framework of Lambda Cold Dark Matter ($\Lambda$CDM)
cosmology, favored by a host of astrophysical data,
Primordial Nucleosynthesis (also known as Big Bang Nucleosythesis-BBN) is considered
as one of the pillars of our understandings of modern cosmology.
In its standard formulation (SBBN) predicts that the amount of light elements (essentially null for
atomic numbers above 7) produced in the first hours after the Planck time depend on
one and only cosmological parameter, the baryon density $\omega_b$, which can nowadays
be retrieved from the refined measurements of the Cosmic Microwave Background
anisotropy spectrum, reading $\omega_b$=0.02249 from WMAP 7 year data, \citet{komatsu11}.
SBBN is therefore a parameter-free theory, whose uncertainties on its final predictions ---the light
element abundances---
depend on the uncertainties on the measured physical quantities which are used. 
In particular, as it is nowadays well established and presented
at this meeting by A. Coc and F. Villante, the reactions necessary in the nuclear
network used in SBBN calculations are well known, thus making the 
systematic uncertainty associated with neglecting potentially relevant reactions virtually
null.
On the other hand, uncertainties associated with the reactions rates
of the relevant reactions do exist, and their estimate is a crucial element for
the evaluation of the error budget of light element abundance predictions;
the (very advanced) state of the art in this field has been presented at this
meeting by A. Coc, and it is extremely satisfactory that estimates performed by
several groups converge within the uncertainties, \citet{coc12}, \citet{serpico04}.

The elemental abundance which permits the best check of SBBN is deuterium:
its extreme sensitivity to the baryon density has given the first historical
test of the standard cosmological model. 
Today, observations of hydrogen rich clouds
absorbing QSO's background light permit accurate observations of virtually
uncontaminated environments, thus measuring what is estimated to be the 
primordially produced deuterium abundance, in light of this isotope's fragility
which leads to its destruction by virtually every process.
As reminded in this meeting by K. Olive, observations
of deuterium abundance in the above mentioned environments, have a 
small spread around a central value, 
{\rm D/H}=2.82$\pm$0.20$\times$10$^{-5}$, \citet{pettini08}.
This is compatible -within the 
uncertainties- with the one predicted by SBBN
{\rm D/H}=2.56$^{+0.11}_{-0.10}\times$10$^{-5}$, \citet{iocco09} rescaled for WMAP7
$\omega_b$ with the \citet{pisanti:2007hk}, as all following theoretical BBN abundances.

Observations of $^4{\rm He}$ in  metal-poor, extragalactic HII regions, with the need for 
a subsequent regression to zero metallicity.
As presented in this meeting by E.~Skillman, it seems that 
after years of uncertainty, and of a bizarre trend of the central value which
increased with the years,
$^4{\rm He}$ observations today converge towards a single value, 
{\rm Y$_p$}=0.2534$\pm$0.0083, \citet{aver12}
This value is broadly compatible with the one predicted by SBBN when using
the very same value of baryon density $\omega_b$ from
CMB observations as above, namely the same SBBN which predicts
Deuterium compatible with observations, {\rm Y$_p$}=0.2467$^{+0.0003}_{-0.0002}$.

The $^7{\rm Li}$ predicted by this very configuration of SBBN is
$X_{\rm Li}$/$H$=4.6$^{+0.2}_{-0.5}\times$10$^{-10}$ , which in the most common 
unit of {\it dex} adopted above reads {\rm A(Li)} = 2.66 $^{+0.02}_{-0.05}$;
this quoted number may vary slightly according to different groups and 
network, and all theoretical estimates are consistent with each other.

Therefore, if the value of  the envelope is to be interpreted as
the primordially produced lithium, there is a friction with SBBN.
Namely, one would be faced with the situation in which SBBN 
is nicely compatible with Deuterium and $^4{\rm He}$ observations,
but is overproducing the $^7{\rm Li}$; in this scenario, a mechanism
is therefore needed in order to reconcile $^7{\rm Li}$ without destroying
the concordance of the other elements.
A. Coc has stated that ``{\it no sensible modification of primordial
$^7{\rm Li}$ is possible through standard BBN}'', and
the possibility that an unknown, experimentally
overlooked resonance in the destruction channels of beryllium-seven
(which makes up approximately half of the $^7{\rm Li}$, by successive decay)
has been described as ``{\it requiring almost unphysical processes}'' by F. Villante
during this meeting, \citet{broggini12}.
Which leaves very little room for justifying the interpretation of the observed $^7{\rm Li}$
as primordial, unless exotic models of BBN -i.e. models with physics beyond the Standard
Model of Particle Physics or Cosmology - are invoked.

The possibility that an astrophysical population (and
its associated phenomena) intervening
between BBN and Population II, may have modified the
lithium abundance in such uniform way is today generally 
discarded as unphysical by the whole community.

\subsection{Exotic BBN models}
The injection of high-energy, non thermal particles during the Big Bang
Nucleosynthesis is known to modify the final abundance of light
primordial elements, as a consequence of non-thermal nuclear reactions,
see e.g. \citet{jedamzik09}.
The entity of the modification to the final light element abundances depends 
quite strongly on the nature and initial energy spectrum of the non-thermal particles injected,
thus making a thorough discussion of such topic unfit for this proceedings.
What is most compelling to report here,
as discussed at the workshop by K.~Olive and T.~Kajino,
is the fact that several Dark Matter candidate models exist that
predict either the annihilation or decay of the Dark Matter particle
into standard model particles  ---or either
the decay of supersymmetric particles into the Dark Matter particle
itself--- during the BBN.
In particular, different results are obtained for the injection of
hadronic showers vs leptonic ones, e.g. .
Additionally, thermal nuclear reactions can be modified if 
negatively charged massive particles (CHAMPs), with a lifetime comparable with
the Hubble time at BBN epoch are present.
Several regions of the SuperSymmetric parameter space (as well
as of other extensions of
the Standard Model),
currently unconstrained by the LHC results as well as from
indirect and direct DM detection, have as a result
to destroy $^7{\rm Li}$ to the level of the plateau roof, see e.g.  \citet{kajino10}.
Many of these models do however also predict an
increased production of primordial $^6{\rm Li}$
---which in SBBN is produced in negligible amounts
$X_{^6{\rm Li}}$=1.1$\times$10$^{-14}\pm$0.1---
which was considered a good feature of such models in light of the 
$^6{\rm Li}$ observations in stellar atmospheres.
The latter seem to be strongly challenged, see following section, 
and in order to obtain only $^7{\rm Li}$ depletion at the plateau level
one must restrict the region of the parameter space where a solution is
available, although the price to pay is almost unavoidably Deuterium
overproduction, see e.g. \citet{olive12}.
Other ways to deplete $^7{\rm Li}$ synthesis during BBN
are phenomenologically viable, including baryon inhomogeneous BBN,
axion cooling mechanisms, variation of constants of nature,
 and have been discussed at this workshop by 
T. Kajino.
%################
\section{Lithium-six, or not?}
\label{sec:lisisx}
%################
$^6{\rm Li}$ detection in stellar atmospheres is extremely challenging:
the isotopic shift makes its feature appear as a small depression in the redward region of the lithium
doublet line.
The precision needed in order to resolve such delicate feature,
%and the accurate knowledge of atmospheres 
had made its detection
impossible until very recently. 
However, the two-sigma detection of $^6{\rm Li}$ in nine stars (or three-sigma detection
in five stars), \citet{asplund06} had stirred much interest.
The abundance X$_{^6Li}\sim$5\% X$_{^7Li}$, seemingly the same in all of the objects of 
different metallicity, was
4 orders of magnitude above the predictions of SBBN, as well as in inexplicable
with models of cosmic ray induced chemical evolution of the Galaxy, as thoroughly
discussed by B.~Fields in his talk, \citet{prodanovic07}.
This circumstance brought to the claim of a $^6{\rm Li}$ ``plateau'', 
ideally analogue to the $^7{\rm Li}$ one, which was soon used to advocate
the primordial nature of the observed $^6{\rm Li}$, the remarkable discrepancy with
SBBN, and the need for exotic models. Many of the latter could self-consistently explain
both the deficit in $^7{\rm Li}$ and the overabundance in $^6{\rm Li}$, as reminded in previous section.
However, as stated during this workshop by A.~Korn: ``{\it it is very unfortunate that we started talking
of a $^6{\rm Li}$ plateau in the first place}''.

In fact, as it has been realized soon after the first claim of detection, the small depression in the red wing of the
lithium doublet line line could be due to effects other than the isotopic shift.
Motions in the atmosphere can easily produce features that in an LTE analysis of the line
are interpreted as evidence for $^6{\rm Li}$, therefore not only 
the sum of the two isotopes abundance and their relative abundance must be adopted
as parameters for the line fitting, but also others describing the line broadening because
of motions of the gas in atmosphere.
In particular, asymmetric flows in the atmosphere cause a line shape distortion comparable with
the one expected by a blend of $^6{\rm Li}$/$^7{\rm Li}$, with an assumed $^7{\rm Li}$ symmetric shape.
Several analysis exploring this scenarios have been carried on:  
K.~Lind and M.~Steffen have devoted their talks to this subject, and details can be found in their
proceedings and references therein.
What it seems compelling to report here is the fact that evidence for detection of
$^6{\rm Li}$ seems to be vanishing, or to put it in other words, $^6{\rm Li}$ can not be 
safely claimed to have been detected in many of the stars of original sample,
and detection is actually excluded for several others, depending on the type of atmospheres adopted.
In particular, according to theoretical corrections based on 3D NLTE predictions of the line profile,
Steffen et al. obtain a reduction from nine to five stars (under the two--sigma criterion) and from 
five to two (under the three sigma criterion). If they adopt results based three-dimensional simulations
of the atmospheres with proper treatment of convection and line formation, they obtain that
a single star (HD 84937) exhibits non-negligible $^6{\rm Li}$ abundance.
For this very same stars, Lind et al. obtain instead a vanishing $^6{\rm Li}$ abundance,
based on an 3D, NLTE analysis of both lithium and Ca line profiles.

Whereas this issue is clearly not yet established, it seems
to go in the direction of telling us that
what has been for few years the`` $^6{\rm Li}$ plateau'', 
is now the measure of a non vanishing $^6{\rm Li}$
abundance in a limited sample of stars.
For as interesting as this can be, at the state of the art this can not be reasonably
considered as an indication for a cosmological signature. 
%################
\section{Conclusions}
%################
`What is the ``Lithium Problem'' today?'
is the relevant question to be asked before looking for its solution.
Namely, what is it about lithium that state-of-the-art observations and theory
fail to explain, and must therefore be regarded as a ``problem''?
Until very few years ago the ``Lithium Problem'' could have been phrased 
this way: ``Observations of $^7{\rm Li}$ in the atmosphere of field halo, turn-off
stars ---spanning almost two orders of magnitude in metallicity--- exhibit a single
value abundance, with virtually zero dispersion around the central (only) value. Such
value is a factor 3 to 5 smaller than what is expected to be leftover by SBBN, and no
known astrophysical mechanism is able to deplete the lithium with such a fine---tuned accuracy.''
Such formulation, would hint toward a solution in terms of either: 
{\it i)} an unknown, fine tuned astrophysical mechanism, at work with same
strength in the atmospheres of very different stars, capable to level the lithium
abundance starting from the SBBN value or
{\it ii)} a modification of SBBN, which would be accurate in predicting all
other light elements but $^7{\rm Li}$, by virtue of wither unknown nuclear processes,
or exotic processes involving non---standard model particles of modifications of the constants of nature.
For both of these explanations, viable physical models existed, however yet unfalsified in
their predictions of other systems.

The observations and analysis of  the last years, discussed at this
workshop, display a somewhat more complicated observational
evidence than only few years ago:

{\it i)} the abundance of $^7{\rm Li}$ in turn-off stars in the halo of the Milky Way 
and in Globular clusters exhibit a ``roof value'' of ${\rm A(Li)}$=2.4 throughout the range of
metallicities [Fe/H]$\leq$-2.0.
Very few outliers can be found above such value, and their number is statistically irrelevant;

{\it ii)} below such envelope value, the observed abundances show:\\
{\it iia)} moderate dispersion of $\sim$0.2 dex in the metallicity range  -2.8$\leq$[Fe/H]$\leq$-2.0\\
{\it iib)} very big dispersion ---down to vanishing abundances--- in the metallicity range [Fe/H]$\leq$-2.8. 

In light of this, the ``Lithium Problem'' today may be rephrased in the following way:
``Observations of $^7{\rm Li}$ in the atmosphere of turn off stars
spanning more than two orders of magnitude in metallicity, do not pass
a roof/envelope abundance a factor $\sim$3 smaller than what predicted from SBBN.
Little, yet non-zero, dispersion is seen below such value at intermediate
metallicities, and much more consistent dispersion -down to vanishing abundances-
is seen from the lowest metallicities.''

This is clearly a much more complicated scenario to be solved with a single answer.
If the explanation is astrophysical, one shall envision a mechanism that from the
SBBN value depletes the lithium of {\it at least} a factor 3 in the atmosphere of metal-poor
stars, with little dispersion (with similar efficiency of the mechanism) at intermediate metallicities,
and with very different efficiency at the lowest metallicities.

On the other hand, if one assumes that some exotic mechanism
is at work during BBN,  such to modify the abundance of the $^7{\rm Li}$ isotope down to the
plateau envelope value, additional problems would still to be faced.
An astrophysical mechanism able to cause strong depletion at low metallicities,
shall be invoked, yet capable of limiting such depletion at the observed level at higher metallicities,
in a mechanism much resembling those that were so fine tuned, and be argued against  
when invoked as solution of the old formulation of the ``Lithium Problem''.
Yet, it seems clear that in both scenarios astrophysical processes
are at work,  even if at low metallicities only, on the top
of an unknown astrophysical, cosmological or particle process,
and that the use of the plateau roof for cosmological use and
constraints on non-standard processes seems riskier today
than it was in the past.

A specular discussion of these arguments can be found
in the proceedings of Sbordone et al. and of  Spite et al.,
and examples of composite explanations have been discussed
at the meeting, see e.g. P.~Molaro's proceedings.

In spite ---or maybe because of--- all the novelties and elements
discussed at the Paris workshop, I find impossible to
write down any Conclusion about the Lithium Problem, and I prefer
to end these proceedings with a somewhat pedantic, yet not inappropriate
semantic thought.\\
Is the Lithium Problem Primordial, Cosmological, Cosmic or Astrophysical:
which of the several adjectives that have been imposed to the problem is the most fitting to its nature?
In light of the previous considerations, I suggest that either Astrophysical,
Cosmological or Primordial would impose on the problem a shade of pre-concept solution,
and that perhaps ``Cosmic'' is the least unfitting definition, given the universality 
of the problem itself with respect to galactic observation loci.

{\it Note added---} It is with much relief that ---after completing this text--- I read the
proceedings of M.~Spite, F.~Spite, and P.~Bonifacio in this volume,
and I realize that a phenomenologist's perspective is after all not
that different from an observer's one.
We leave it to the reader to investigate the consequences of this worrisome finding.
% for as worrying as this may seem at a first glance.

\begin{acknowledgements}
I would like to thank all the participants of the
\href{http://www.iap.fr/lithiuminthecosmos2012}{{\it Lithium in the Cosmos}} 
workshop for making it the enriching meeting it has been,
and all the LOC and SOC members for the hard work which has made it
possible. A specially warm thank goes to E.~Vangioni for
her continuous support in both scientific and practical
matters throughout all the phases of the organization.
A.~Cuoco and P.D.~Serpico are kindly acknowledged for
proofreading, and O.~Pisanti for providing the WMAP7 cosmology
runs of the \href{http://parthenope.na.infn.it/}{PArthENoPE} code.
\end{acknowledgements}

\end{document}